# Synthesis and *in-situ* characterization of photochromic yttrium oxyhydride grown by reactive e⁻-beam evaporation

K. Kantre[1,*], M. V. Moro[1], D. Moldarev[1,2], M. Wolff[1] and D. Primetzhofer[1]

[1] Department of Physics and Astronomy, Uppsala University, Box 516, S-751 20 Uppsala, Sweden
[2] Department of Material Science, Moscow Engineering Physics Institute, Moscow, 115409, Russia

**Abstract**

We report on controlled growth of photochromic yttrium oxyhydride thin films monitored by *in-situ* composition depth profiling. Films were grown by reactive e⁻-beam evaporation and subsequently oxidized, while simultaneously tracking the oxygen and hydrogen concentrations. Sample composition and photochromic response were characterized *in-situ* using non-destructive ion beam analysis and image analysis, respectively – as well as complementary *ex-situ* ion beam methods, X-ray diffraction and optical spectrophotometry. We show that photochromic yttrium oxyhydride can be grown as yttrium dihydride, which is then oxidized to O/H ratios triggering the photochromic response.

*Corresponding author: karim.kantre@physics.uu.se

Photochromic materials offer exciting applications as smart window materials with potential global impact on energy consumption by reducing heating and cooling costs via passive regulation of heat flow into and out of buildings [1,2]. Yttrium oxyhydride ($YH_xO_y$) thin films have recently attracted the attention of the scientific community as they can exhibit reversible photochromism at ambient conditions [3]. By changing deposition parameters and thus chemical composition, the optical band gap [4] and electrical resistivity [5] of the film can be tuned and it was shown that the photochromic response strongly depends on the oxygen/hydrogen content [6]. More recently, a similar behavior was also observed for rare-earth metal oxyhydrides (ScHO, NdHO, GdHO, DyHO and ErHO) [7–9].

Yttrium oxyhydrides are usually grown by reactive sputtering of yttrium in $Ar/H_2$ atmosphere and post oxidation in air [10]. In the very first studies the origin of oxygen in the films was unclear and it was assumed that it is incorporated during growth [3]. However, more recent investigations indicated that oxygen is incorporated after the sample is exposed to ambient conditions [11]. Stability studies carried out on $YH_xO_y$ samples, have shown that film oxidation can continue slowly in air over periods of days to months [12]. Additionally, it was found that film deposition by magnetron sputtering favors formation of columnar domains that may enhance oxidation [9].

A precise control and monitoring of the composition during deposition is yet unavailable. Questions on how oxidation rate and details of the selected deposition method (with the resulting film structure) affect magnitude, switching rate and stability of the photochromism observed are yet to be answered.

In order to get a better handle on the synthesis of photochromic yttrium oxyhydride *in-situ* chemical composition analysis and optical characterization during the actual growth are desirable. We present composition analysis by non-destructive Ion Beam Analysis (IBA) [13] during sample synthesis and combine the results with *in-situ* optical measurements.

We synthesized photochromic $YH_xO_y$ thin films, for the first time, by a combination of reactive e$^-$-beam evaporation and controlled post-oxidation. After synthesis samples were characterized with respect to the chemical composition and photochromic effect *in-situ/in-vacuo*. Complementary information on composition, structure and optical properties were subsequently obtained *ex-situ*, after exposing the samples to air, including further chemical composition analysis by ion beams.

The *in-situ* investigations were performed in SIGMA (Set-up for *In-situ* Growth, Material modification and Analysis) [14], located at the 5 MV Tandem accelerator at Uppsala University [15]. The experimental set-up has the capability of studies of material growth, such as inert and reactive thin film deposition, thermal treatments and ion implantation (i.e. material modification) combined with *in-situ* IBA characterization. Films were synthesized in ultra-high vacuum, while simultaneously the oxygen and hydrogen concentration, partially depth profiled, were measured *in-situ* by Rutherford/Elastic Backscattering Spectrometry (RBS/EBS) and Elastic Recoil Detection Analysis (ERDA) after each process step. Additionally, Particle-Induced X-Ray Emission (PIXE) was performed employing the same beam. Optical characterization of the photochromic response was performed both *in-situ* and *ex-situ*. The composition of samples was measured *ex-situ* by Time-of-Flight ERDA (ToF-ERDA) and structural analysis was done by X-ray Diffraction (XRD).

The films were deposited on transparent soda-lime glass (microscope slides, 20x10 mm$^2$ and 1 mm thick) and polycrystalline $CaF_2$ (12 mm diameter and 1 mm thick) substrates allowing measurements of optical transmission. Yttrium hydride ($YH_x$) thin films were grown by reactive e$^-$-beam evaporation of a commercial metallic yttrium target (99.9% of nominal purity) from a tungsten crucible in a reactive hydrogen atmosphere (kept constant at $3\times10^{-4}$ Pa) using a triple source e$^-$-beam evaporator (FOCUS EFM 3T). The substrates were loaded onto a goniometer at 100 mm distance from the evaporator. The yttrium target was degassed prior to evaporation and the base pressure in the chamber before growth was lower than $5\times10^{-7}$ Pa. Then oxygen was introduced in the chamber by a controlled flux via a leak valve, with doses up to $10^7$ Langmuir to oxidize the film. In an alternative oxidation path, samples have been oxidized by direct exposure to atmospheric conditions, similarly to the synthesis process usually followed for the growth of photochromic $YH_xO_y$ by reactive magnetron sputtering.

The experimental geometry for reactive $YH_xO_y$ growth, *in-situ* IBA characterization and *in-vacuo* characterization of photochromism are illustrated in Figure 1a. Samples were facing the e⁻-beam evaporator during deposition and were rotated for simultaneous RBS/EBS, PIXE and ERDA measurements (rotation indicated by a dashed arrow in Figure 1a). The angle between the ion beam direction and sample surface normal is 65.5°, allowing simultaneously RBS/EBS and ERDA measurements. For oxygen detection, we performed EBS with the detector placed at a scattering angle of 170° and using the narrow (≈ 10 keV) elastic resonance of $^{16}O(α,α_0)^{16}O$ at 3037 keV [16]. The amount of yttrium and the ratio of both elements (i.e., O and Y - see left panel in Figure 1b) were also extracted from these measurements. To reveal the hydrogen concentrations, ERDA was performed at a detection angle of 27° with respect to the beam direction. Hydrogen recoils from the sample were detected, while scattered primary ions as well as other recoiling species were stopped by an Al absorber foil with a thickness of 12.9 μm placed in front of the detector. A 3060 keV He⁺ beam was used to enhance the oxygen detection deeper in the film and avoid the oxide layer close to the surface [17]. Typical ERDA spectra from a $YH_xO_y$ film before and after controlled oxidation are shown in Figure 1b (middle panel). In this geometry, hydrogen recoils have a maximum energy of 1561 keV. An X-ray detector with 204 μm Mylar absorber is placed at 135° scattering angle (see Figure 1a) to identify heavier constituents, indicate possible presence of trace elements in the film and to be used for spectra normalization. Typical PIXE spectra recorded simultaneously with EBS/ERDA are shown in Figure 1b, right panel. Characteristic X-rays from both Y in the film and Ca, in the substrate, which are only insignificantly absorbed in the grown films are visible in the spectra.

The ERDA spectra were normalized by the charge-solid angle product, using the substrate signal in the corresponding RBS spectra (see left panel in Figure 1b). The hydrogen content derived by ERDA is used as a boundary condition to fit the RBS/EBS spectrum in an iterative and self-consistent approach [18]. Even though there is no direct signal from hydrogen in the RBS/EBS spectrum, differences in $YH_xO_y$ composition affect the stopping power in the film and thus the shape of the yttrium peak for constant areal density of yttrium (a comparison of spectra obtained from pre- and post- oxidized films is shown in Figure 1b). The RBS/EBS and ERDA fits were performed by SIMNRA [19], using the latest version of SRIM [20] as a default choice for input values of stopping power. In the case of $^{16}O(\alpha,\alpha_0)^{16}O$ and $^1H(\alpha,p)^4He$ elastic scattering and recoil cross sections, we used data evaluated by SigmaCalc [21].

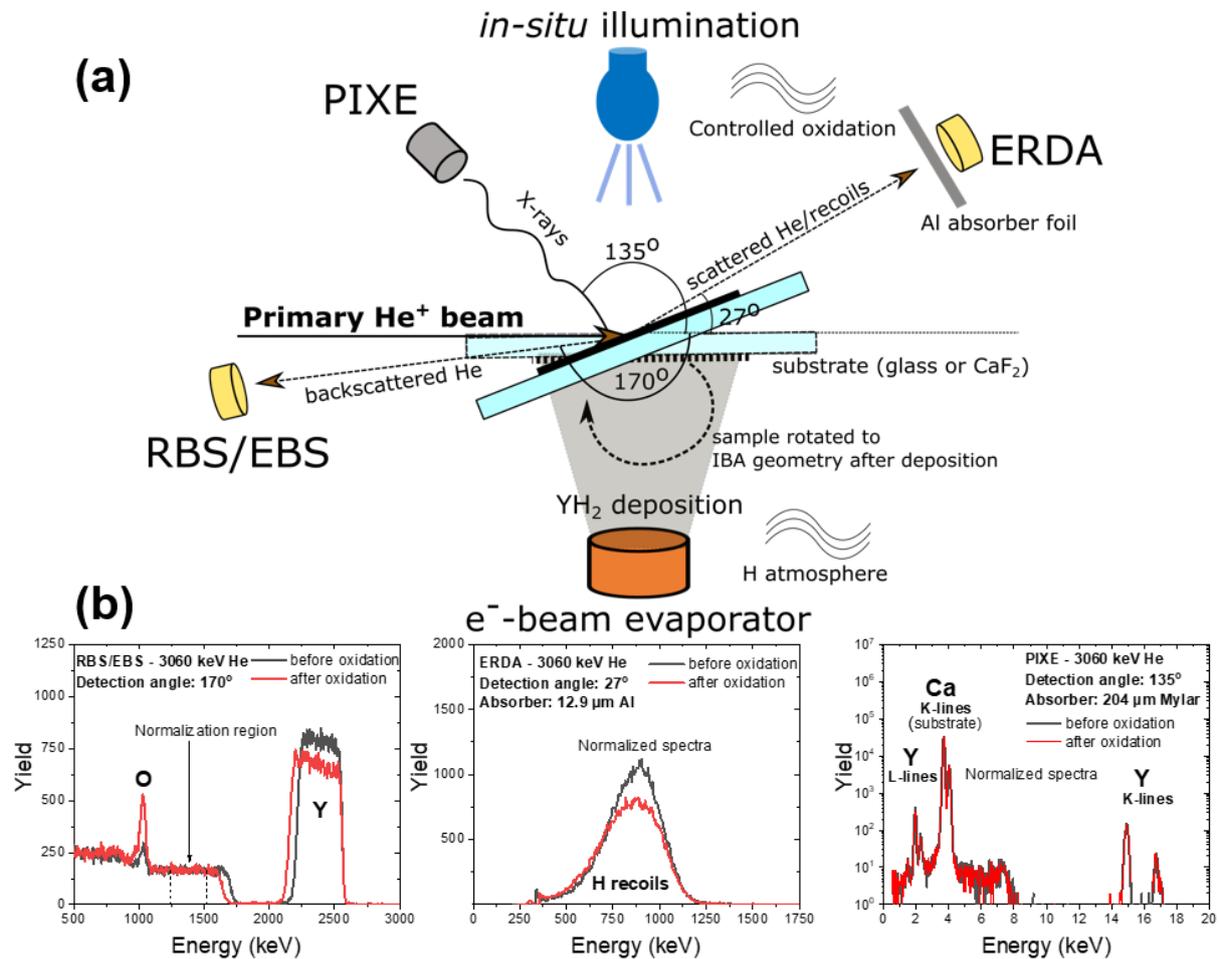

*Figure 1. (a) Sketch of the experimental geometries and set-up for the in-situ studies. Yttrium dihydride is deposited and then $YH_xO_y$ is synthesized in-situ by controlled oxidation. Oxygen and hydrogen are monitored by ion beam analysis. (b) From left to right: RBS/EBS, ERDA and PIXE spectra simultaneously acquired in-situ before (black lines) and after (red lines) oxidation of a thin film deposited on $CaF_2$. The spectra are normalized by current by using the substrate signal indicated in the RBS/EBS spectrum*

To characterize the photochromic response *in-situ* samples were illuminated by a LED-cluster (λ = 455 nm and power = 4.865 W) in IBA geometry (Figure 1a). The light intensity on the sample was enhanced by using a focusing lens (focal length = 14 cm). The induced change in the color of the samples was tracked by repeatedly taking images under identical conditions, using a digital camera (*Canon model EOS4000D*).

The chemical composition was additionally measured *ex-situ* using the ToF-ERDA set-up at the Tandem laboratory [15] to investigate post-oxidation of the samples after air exposure. A primary beam of 36 MeV $I^{8+}$ was employed. Recoils were detected by a ToF-Energy detection system, consisting of two carbon foils measuring the time of flight followed by an ionization chamber for the energy signal [22]. The detector is placed at an angle of 45º with respect to the beam direction. The data were evaluated and depth profiles were obtained by Potku [23]. Systematic uncertainties occurring due to the Time-of-Flight detection efficiency of light elements [24] were taken into account.

Additional optical characterization and structural analysis were performed *ex-situ* by optical spectrophotometry and grazing incidence XRD, respectively (see supplementary information for further details).

After e$^-$-beam growth, all films (resulting nominal thickness from 100 nm to 1 μm) were found to be black, with the color shifting with oxidation to yellow-transparent and finally to fully transparent. Our samples can be divided into three categories with respect to the specific oxidation and characterization procedure. A subset of samples was taken out from the chamber after growth, oxidized in air and characterized *ex-situ*. A second subset was first oxidized *in-situ* before it was exposed to atmosphere and characterized *ex-situ*. For the final subset we followed the full *in-situ* procedure as described above (i.e. growth, oxidation, compositional and optical characterization performed in SIGMA). Eventually, these samples were also characterized *ex-situ* after exposure to air. The measured compositions of samples from all subsets are plotted in the ternary plot of Figure 2 and compared to data from literature [6] of samples grown by magnetron sputtering (open squares).

Samples oxidized by uncontrolled exposure to air are noted by green numerical notation (subset 1). All samples of this subset exhibited reversible photochromism in ambient conditions. Optical transmission spectra exemplifying the photochromic effect are shown in Figure S1 in the supplementing materials. All samples investigated exhibited color neutral photochromic response. The *ex-situ* ToF-ERDA data plotted in Figure 2 were obtained three to nine days after synthesis, while the films were still photochromic. Prolonged post oxidation of the samples in air affects the color of the films, as it shifts to fully transparent, as well as the strength of the effect, as the photochromic properties gradually disappear.

Samples of the second subset are noted by red numerical notation (subset 2) in Figure 2. These samples did not exhibit any (or, in the case of sample 3, very weak) photochromism after they were exposed to air. They were measured *ex-situ* by ToF-ERDA one to seven days after growth. All of them were oxidized *in-situ* and in the case of sample 4 at elevated temperature (150 ºC), which led to absorption of significantly more oxygen.

Blue squares in Figure 2 show the evolution of the composition of one sample throughout its *in-situ* investigation. Initially, after growth, the color of the films was black and its composition was close to that of $YH_2$. At this stage the oxygen content was found to be 6.5 at.%. As the film got oxidized (dose = $9 \times 10^6$ Langmuir), it became yellow-transparent and its composition approximately followed the chemical formula $YH_{2-\delta}O_\delta$ (as suggested by *Moldarev et al.* [6] for films grown by sputtering). In this formula, δ represents the oxygen/yttrium ratio. At δ=0.84 the photochromic effect was triggered *in-situ*. The composition of the sample was not affected by photodarkening. When the sample was exposed to air its color shifted to fully transparent and no photochromism was observed. This behavior is in accordance to what was stated above for samples from the second subset. *Ex-situ* characterization by ToF-ERDA showed that the sample reached a stoichiometry of $Y_{0.26}H_{0.17}O_{0.57}$ after 16 days of post-oxidation in air. While this composition deviates significantly from the proposed formula, the difference can be attributed to water incorporation in the film. The blue dashed line in Figure 2 indicates the expected change in composition when removing water molecules from the film. Higher water absorption as compared to sputtered film is expected due to lower density of e⁻-deposited samples [25].

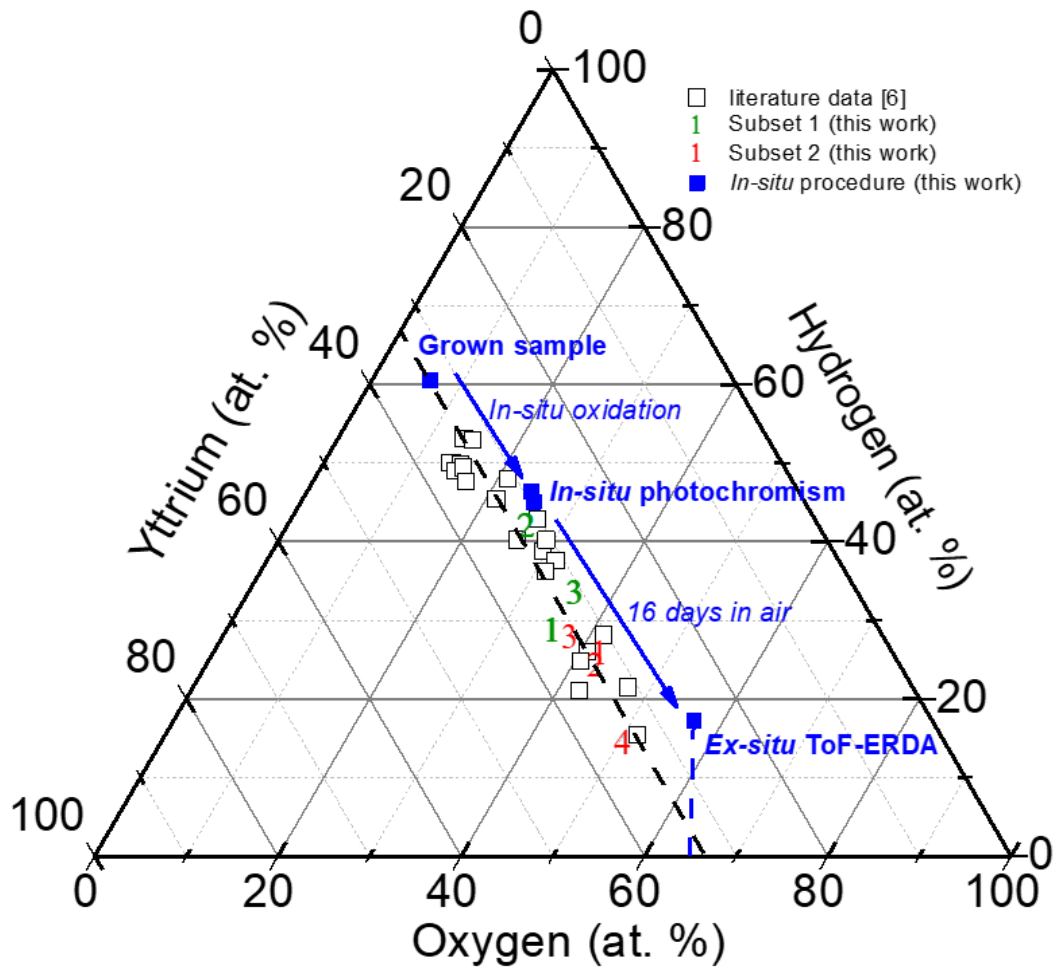

*Figure 2. Composition of samples grown by e⁻-beam evaporation, oxidized by two different procedures and measured ex-situ (green and red numerical notation) plotted together with the composition evolution of a sample followed the fully in-situ procedure in SIGMA (blue squares). Results are compared to samples grown by magnetron sputtering (open squares) [6].*

In Figure 3, photographic images of a film in vacuum, before and after oxidation and after illumination, together with its *in-situ* optical characterization are shown. The brightness profile was obtained across the green line using *ImageJ* [26]. The bare glass substrate regions (indicated by B in the photos) were used to normalize the transmission of the film and compare colors. A comparison is shown between the brightness profiles of the film before illumination and after 8 and 60 minutes of illumination clearly confirming the photochromic properties of the film.

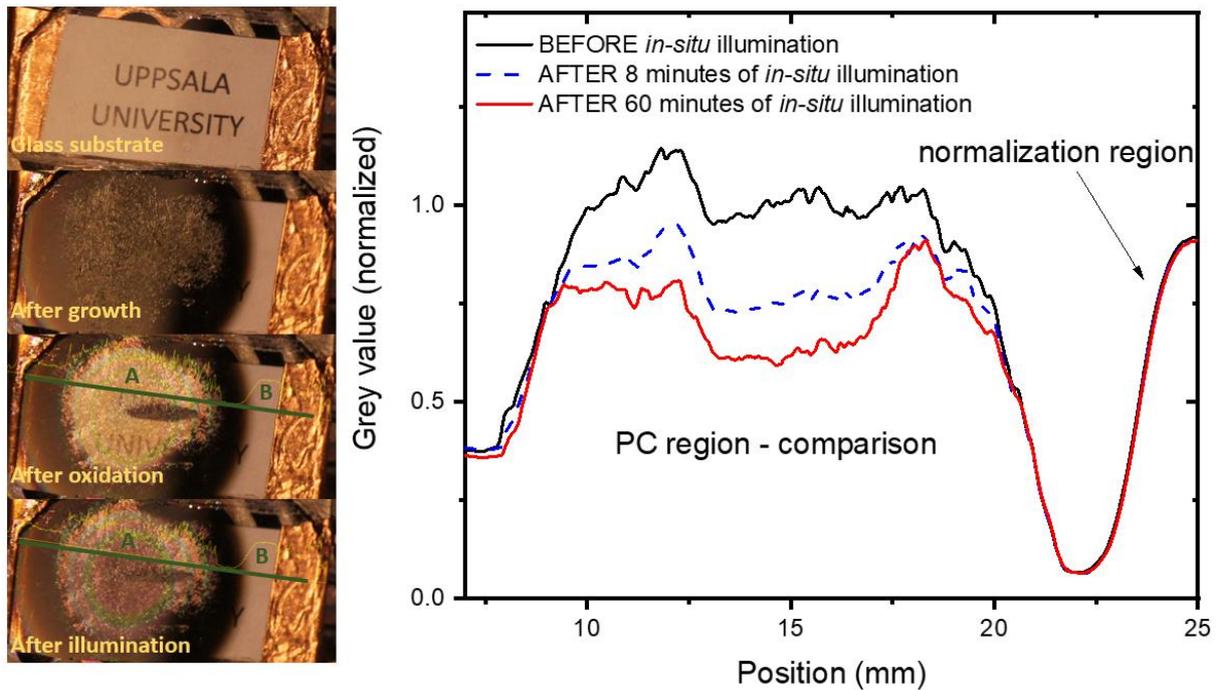

*Figure 3. In-situ optical analysis of a $YH_xO_y$ film synthesized in SIGMA. Left: images taken before growth, before oxidation and before and after illumination. The green line shows the region at which intensity profiles (right panel) were obtained. Photochromic and normalization regions are indicated by A, and B, respectively.*

The behavior described above for all 3 subsets was observed for samples grown both on $CaF_2$ and on glass, i.e. no substrate-dependent effects were found. The results indicate equal properties of e⁻-beam grown and magnetron sputtered films, regarding the correlation of the photochromic response with sample composition, as e⁻-beam grown $YH_xO_y$ films agree in composition with what was proposed by *Moldarev et al.* [6] and exhibit photochromic properties at similar stoichiometry.

Quick oxidation (i.e. by direct exposure of the sample to atmosphere after the deposition of $YH_2$) favors formation of more stable yttrium oxyhydride films that maintain their photochromic properties in ambient conditions for a longer time. As an example in [12], a photochromic behavior in sputtered $YH_xO_y$ samples over seven months was reported, whereas in our case, the photochromic contrast amongst e⁻-beam samples vanished within a few hours to days after air exposure. Samples that are slowly oxidized in the vacuum system tend to have a significantly higher post oxidation rate after their exposure to ambient conditions. This effect might be explained by a changing microstructure of the film under fast oxidation hindering further oxygen incorporation.

In both scenarios, the post-oxidation rate is higher for e⁻-beam grown samples compared to those produced by reactive magnetron sputtering followed by an uncontrolled post-oxidation process. To summarize, while the deposition method is not a decisive factor for photochromism of yttrium oxyhydride films, it is found that it affects the persistence of the effect, since the different microstructure formed by the precipitation of the evaporated yttrium and the resulting lower density, might allow faster oxidation.

Figure S2 in the supplementary material shows the XRD pattern recorded for an *in-situ* oxidized $YH_xO_y$ film grown on glass. Four Bragg peaks as well as the signal of the amorphous substrate are resolved. Similar to sputtered films [27], our sample has a $YH_2$ (FCC) crystal structure [28] with expanded unit cell under oxygen incorporation. The lattice parameter is a=5.37 Å. In recent studies [5], the lattice parameter was found to be between 5.28 and 5.36 Å. The large lattice parameter in our case can be attributed to the high oxygen concentration of the investigated sample.

To summarize, we produced photochromic yttrium oxyhydrides, for the first time by reactive e⁻-beam evaporation and controlled oxidation. Samples were characterized *in-situ* and *ex-situ*. The composition of the thin films was tracked during synthesis *in-situ* by ion beam analysis and the photochromic response was triggered in vacuum. Slow oxidation after synthesis results in faster post-oxidation rates and loss of photochromic effect, which is related to different microstructures. The ability of *in-situ* growth and characterization, combining a well-controlled synthesis and triggering photochromic properties of yttrium oxyhydride thin-films, offers new pathways for optimizing materials properties. It also unlocks the study of different oxidation processes and a more general study of diffusion properties via isotope-labeled metal-oxyhydrides.

The authors acknowledge financial support from the Swedish research council, VR-RFI (contracts #821-2012-5144 & #2017-00646_9), and the Swedish Foundation for Strategic Research (SSF, contract RIF14-0053) supporting the operation of the accelerator and the construction of SIGMA.